\documentstyle[12pt]{article}
\begin{document}
\begin{center}

{\large {\bf On the Unnikrishnan approach to the notion of locality}}

\vspace{2mm}
{\bf Bernard d'Espagnat}

\vspace{2mm}
{\it Abstract}

\vspace{2mm}
{\it Recent proposals by C.S. Unnikrishnan concerning locality and Bell's
theorem are critically analysed.}
\end{center}   

\noindent{\bf 1 - Introduction}

In what follows, two particulars play a major role. One of them (i) has to do with the quantum
 mechanical von Neumann formalism. It is that, according to the latter, not only wavefunctions get 
reduced upon measurement but even state reductions at space-like separated points seem to intervene 
in some cases; specially in those involving measurements on entangled systems. The other, quite 
different, one (ii) is that knowledge of elementary classical physics gets students (including 
future physicists) used to the idea that all of the basic entities science considers should be 
ontologically real. That fields, in particular, do exist by themselves, with the properties we 
ascribe to them, quite independently of us; or, at least, that assuming they enjoy such an 
existence and such properties is often useful and always harmless. This has led many beginners 
in quantum physics to take it for granted that, being quite basic, the wavefunctions necessarily 
have such an ontological reality. 

	The conjunction of facts (i) and (ii) results in that occurrence, in the formalism, of state
reductions at a distance is naturally viewed by students in the field as the clearest evidence
 we may have in favor of nonlocality. And this, in turn, may prompt a theorist inclined to 
question the pertinence of this notion to consider that the task of restoring locality should 
primarily be a matter of disproving state reductions at a distance. It may well be that this 
kind of approach to the subject was that of Unnikrishnan. Indeed, this author recently published 
a series of papers [1],[2],[3],[4] concerning nonlocality and Bell's theorem and, for instance, 
in the abstract of [4], with regard to state reduction at space-like separated points, he wrote:
 ``I show conclusively that there is no such spooky state reduction, vindicating the strong views 
against nonlocality held by Einstein and Popper''. And it is only after having made this statement 
concerning wavefunction reduction that he added: ``The Bell's inequalities arise due to ignoring 
the phase information in the correlation function and not due to nonlocality''.  This, together 
with several other statements of the same vein in his papers, is an indication that indeed his 
reflection proceeded {\it from} a critical analysis of the wavefunction reduction notion {\it to}
 a criticism of the Bell theorem, rather that the opposite way.
	
	Admittedly there are reasons that, a priori, make such a march of thought quite normal and 
natural. It is, in particular, strictly true that, historically, the presence, in von Neumann's 
formalism, of such ``spooky state reductions'' was the first hint pointing to the idea that, within
 physics, the notion of locality might conceivably be questioned. So that it is fully 
understandable that Unnikrishnan should have had his attention first drawn on them. This 
- let it be stressed here - had a positive overall effect since the question he thus focused on,
 that of knowing whether the notion of wavefunction collapse at a distance is contradiction-free 
or not, even though it is an old one, still seems, at present, not completely elucidated. His 
pointing out of this has rightly attracted attention.
	
	Because of point (ii) above, it is also understandable that, having finally reached the 
conclusion that such wavefunction reductions at a distance do not occur, Unnikrishnan 
interpreted it as vindicating Einstein locality. And that this prompted him to look for 
loopholes in the Bell proof of nonlocality. It is, to repeat, understandable and 
natural. And it is therefore most unfortunate that it happened such a rational move 
led him astray. The reason why it did is that, as defined by Einstein (and Bell) the 
notion of locality cannot be dissociated from that of elements of reality, taken in the 
sense of ``elements that really exist, quite independently of our knowledge''. So that, 
with this Einsteinian definition of the words, the notion of wavefunction reduction at 
a distance can possibly have an impact on the locality problem only if the wavefunction 
is considered as being a real entity. For, obviously, under the opposite assumption - the 
view that the wavefunction merely describes our knowledge - the idea that, upon some 
measurements, it gets reduced at a distance, even if true, merely concerns our {\it knowledge}. It 
has no bearing whatsoever on the existence or nonexistence of a real, physical event at the 
place where the reduction is considered as occurring. Now there are sensible arguments in 
favor of such an ``operationalist'' conception. One of them is the fact that the idea the wave
function of a physical system is ``real stuff'' is in no way an integral part of the quantum 
mechanical formalism. At any rate, it lies much beyond the ``hard core'' of what the latter 
consists of, which essentially is a set of general (and never falsified) general rules 
yielding the probabilities of such and such {\bf observations}. Another, even more weighty, 
argument is that, conceptually, the idea that the wave function is real stuff is - obviously - 
quite difficult to reconcile with the general notion of measurement collapse.  Indeed, it seems 
that the latter argument dissuades most physicists - including those with a ``realistic'' turn 
of mind - from explicitly referring, in their reasoning, to the reality of the wave 
function. This is why, a fortiori, on the tricky question whether or not reduction at a 
distance is real or not, most of them take up an attitude of reserve. 
	
	Things being so, it must be granted that what, concerning the wavefunction, we know for 
sure is merely that it is an efficient tool for predicting what will be observed. It 
follows from this that - contrary to what Unnikrishnan's thinks is true - we shall be 
justified in dissociating the conceptual problems concerning reality from - as it 
turns out - more ``technical'' ones, concerning quantum mechanics. To claim - as Unnikrishnan 
did - that specific notions (such as phases and so on) associated with wavefunctions must 
necessarily be introduced in analyses of questions concerning elements of reality and 
Einstein locality seems quite natural at first sight but is in fact preposterous. It may 
quite well happen - and historically it did often happen - that concepts and 
representations that are most useful for some purposes get nevertheless superseded, 
when more general purposes are at stake, by entirely different ones. When considering 
one particular problem it is therefore advisable to - whenever possible - carry on the 
relevant investigations without assuming at the start the absolute validity of some 
currently used notions.  And this, of course, is especially true in cases in which, 
as here, some obstacles exist, making it objectionable to raise the notions in question 
to the level of true descriptions of what ``really is''.
	
	Here we choose therefore to exclusively discuss Unnikrishnan's  analysis of locality and 
the Bell theorem. This implies that the criticisms we shall state exclusively bear on 
his treatment of this subject. To repeat, his questioning concerning collapses at a 
distance is interesting in itself, independently of any specific interpretation (realistic 
or otherwise) of the quantum mechanical rules, Simply: it will not be investigated in 
what follows. 
	
	The criticisms in question are explained in Section 2, which constitutes the essential 
part of the present paper. Sections 3 to 6 are to be viewed as answering possible objections 
and bringing in additional information.

\vspace{5mm}
\noindent{\bf 2 - Locality and Bell's theorem.}

	Bell of course acknowledged the tremendous power of quantum mechanics at predicting future 
observations on the basis of past knowledge. But he did not rest content with just this. He 
was a realist in the sense that he considered physicists should strive at knowing reality as 
it really is, quite independently of us. And he was clear-sighted enough to realize that 
quantum mechanics meets with hosts of difficulties when it is interpreted as yielding a 
knowledge of such a kind. Concerning it he therefore held the somewhat ``operationalist'' 
view sketched above. He  regarded it as being essentially - at least in its present stage - 
just a set of efficient previsional recipes, with the consequence that he did not attach 
any binding {\it ontological} significance to the quantum mechanical concepts. What he therefore 
tried to do - and succeeeded in doing - was to derive, concerning reality per se, some 
(possibly negative) knowledge independent of the said concepts and directly grounded on 
experiment. Let it be stressed once more that Unnikrishnan's standpoint is quite at 
variance with this. For example, in section 1.4 of [2] - entitled: ``Where is the phase'' - 
he insisted that ``every reasonable description of microscopic phenomena should have 
theoretical 
constructs at the single particle level that can represent the wave nature, and specially 
the phase''. Now, as pointed out above, strictly speaking, this statement of his is not true. 
It is too conservative. The ``should'' that appears in it indicates a conception of quantum 
mechanics that unduly raises to the level of elements that {\it must} be taken into account 
notions (here the phases) that may well just be parts of a successful predictive recipe. 
As if, in Astronomy, noting that, notwithstanding its descriptive deficiency, the geocentric 
theory can serve as a recipe for predicting observations - eclipses, say -, we claimed: 
``therefore, any reasonable description of the Solar System should have theoretical constructs
of the nature of epicycles''. This constitutes my first criticism to his approach.

	My second criticism reaches even more to the crux of the matter. It is that Unnikrishnan 
completely missed the nature and bearing of Bell's theorem. To repeat: Bell did not consider 
the ``quantum states'' as being objective, that is as being the states in which elements of 
``reality per se'' really are. But of course, being a realist, he nevertheless had to consider 
``objective states'', that is, states in which physical systems can {\it be}. Concerning the nature 
of these objective states he kept extremely general. In fact, he merely postulated that 
they are specified by a (discrete or continuous) set of parameters {\bf h} (commonly called 
``hidden'') having the nature of real numbers. He then - still quite generally - observed 
that even spacelike-separated events may exhibit correlations since they can be influenced 
by past common causes, but that it is natural to believe that, when they do, it is exclusively 
for this reason. And that, consequently, within subensembles in which the common causes are 
fixed, no correlation should take place. This natural idea is what he called ``local 
causality'' or, sometimes, just ``locality''. An often overlooked but nevertheless crucial 
point is that it is {\it merely from these general views} that he could derive, for the 
correlation function P({\bf a},{\bf b}), an expression leading to predictions that are violated by 
the experimental data, thus disproving local causality. Therefore Unnikrishnan's statement 
(in Section 1.4 of [2]) that ``there is no quantity in  P({\bf a},{\bf b}) that reflects the phase 
information associated with the individual particle'' is just simply false. According to 
Bell, the parameters {\bf h} are, by definition, {\it all} the parameters (known or unknown, knowable 
or unknowable) that define the objective state of the pair. And he could prove his theorem
without assuming anything concerning the nature of these parameters (except that they are 
real numbers). Consequently, if it is considered that there are ``phases'' among them (as would
have to be assumed by a physicist inclined to interpret quantum mechanics in an ontological way), 
these phases are ipso facto included within the set {\bf h}. As we see, one of the main interests of 
Bell's proof lies in its generality. It holds good whatever the structure is of the physical 
theory that we care to believe in, and supposes a meaning of the word ``locality'' extending 
but minimally beyond the one in which it is commonly used 
	
	To this quite basic criticism let a few remarks be added, that illustrate the extent to 
which Unnikrishnan misunderstood Bell's position. First, let it be noted that, while 
Equation (5) of [2] reproduces with but minor changes Bell's correlation function P({\bf a},
{\bf b}), 
the there accompanying specifications of the nature of the {\bf  h}'s and of A and B are faulty. 
{\bf  h} (the distinction between {\bf  h}$_1$ and {\bf  h}$_2$ is inappropriate) indeed collectively designates a 
set of hidden variables but, contrary to Unnikrishnan's assertion, these variables are 
not associated with the {\it  outcomes}. They characterize (completely) the objective state of 
the particle pair before the measurements take place (otherwise said, the ``common causes'' 
at the source). And as for P({\bf a},{\bf b}) itself, it is an average, but not on ``the product of 
eigenvalues'' as stated by Unnikrishnan. Indeed, since, as we saw, quantum mechanics is 
not involved in this derivation, the notion of eigenvalues is here radically irrelevant. 
And, for the same reason, Unnikrishnan's statement in Section 2.1 that ``the defect in the 
local realistic theories is that the correlation is calculated essentially by averaging 
over the product of eigenvalues'' is meaningless. In the version of Bell's theorem for 
which formula (5) of [2] holds true, determinism is assumed and A({\bf a},{\bf h}) and B({\bf b},
{\bf h}) are, 
respectively, the outcomes (+ or -) that are bound to emerge as results of measurements 
performed with instrument settings {\bf a} and {\bf b}, when the parameters of the pair are {\bf h}. In view of 
all this it is quite clear that Unnikrishnan's statement: ``The assumptions that went into 
the formulation of Bell's theorem have not encompassed the essential physical properties'' 
([2], Section 1.4) is plainly erroneous. It just stems from the fact that Unnikrishnan did 
not understand what Bell did.

\vspace{5mm}
\noindent{\bf 3 -Unnikrishnan's model is not a counter-example}

	In the past, many attempts were made at building up counter-examples to Bell's theorem, 
with the aim of proving that the latter relies on some additional implicit assumptions. 
All of them failed. But a priori it might be conjectured that the model Unnikrishnan describes 
in [2], Section 2, is a new, and (this time) successful, counter-example. Indeed, this model 
does reproduce the quantum mechanical predictions and Unnikrishnan puts forward arguments to 
the effect of showing that it is local. So, does its existence show Bell's theorem is flawed? 
In fact, it is easy to see that the answer is ``no'', for the author's considerations are based 
on a definition of locality that is totally different from Bell's one.
  
\vspace{5mm} 

\noindent{\bf 4 - The definitions of locality: two reasonable candidates}

	The foregoing remark should induce us to have a look at various possible definitions of 
locality and try to appreciate their relative pertinacy. Clearly, since we want to keep general 
we must not postulate determinism. In considering indeterminist theories we cannot, of course, 
assume that the outcomes A and B of the measurements are {\it functions} of {\bf a} and {\bf h} and 
{\bf b} and {\bf h} 
respectively (in the above used notations). Hence, we must generalize the ``local causality'' 
notion. This Bell and others did as follows. Denoting quite generally by the symbol (X$\vert$Y) 
the conditional probability that ``X if Y'', they identified ``local causality'' (alias, simply 
``locality'') with the assumption - call it ASN - that the conditional probability (A$\vert${\bf a},
{\bf h}) 
that ``the outcome on the first instrument is A when the setting of the instrument is {\bf a} and 
the hidden parameters are {\bf h}'' is independent of both b {\it and} B; and symmetrically, of course, 
(interchanging A and B as well as a and b) concerning the second instrument. For the above 
stated reason, such an assumption is a priori quite a natural one, adequately reflecting what 
we intuitively have in mind when we assert that, within two, distant, given ensembles of random 
events, the events in any one of the ensembles exert no influence on those of the other ensemble. 
From the general rules of probability calculus the joint conditional probability that ``the 
outcomes on the first and second instruments are A and B respectively, given that the settings 
are {\bf a} and {\bf b} and the hidden parameters of the pair are {\bf h}'' is 
(A$\vert${\bf a},{\bf b},B,{\bf h})(B$\vert${\bf a},{\bf b},{\bf h}) and because of assumption ASN 
this reduces to just the product 
(A$\vert${\bf a},{\bf h)(B$\vert$b},{\bf h}), so that the correlation function P({\bf a},{\bf b}) 
is of the form:

$$	P({\bf a},{\bf b}) = \int d{\bf h} \rho({\bf h})(A\vert {\bf a},{\bf h})(B\vert {\bf b},{\bf h})\ \ \  \mbox{(I)}$$
where $\rho$({\bf h}) is the statistical distribution of the hidden parameters. And the 
Bell theorem in its ``final'' form consists in the proof that both the quantum mechanical 
predictions and the experimental data violate this relation (I), that is, violate the, 
thus generalized, ``locality''. 

	We have just seen Bell's definition of this notion. But we must remember that the way 
we associate words with concepts is, to some extent, a matter of free choice, and this 
is particularly true when, as here, the word in question is an everyday one and we need 
use it outside the commonsense domain. It is therefore not surprising that, in the 
relevant literature, at least two reasonable definitions of the word ``locality'' were 
put forward, with the (admittedly unpleasant!) consequence that the same set of data 
is compatible with locality when the word is defined one way and is not when the word 
is defined in the other way. More precisely, when defining ``locality'' above, we specified, 
along Bell's lines, that locality means the conditional probability of event A for fixed 
{\bf a} and {\bf h}'s is independent of both {\bf b} and B (and conversely, exchanging A and B 
as well as {\bf a} 
and {\bf b}). Some physicists opted for a different choice. They decided to consider that locality 
holds true as soon as the conditional probability in question is independent of {\bf b}, 
no condition 
being specified involving B. Such a weak definition of ``locality'' is also called ``parameter
independence''. It suffices to guarantee that no supraluminal signaling can take place. And 
it could be shown that, with it, no theorem similar to Bell's one can be proven, so that 
models can easily be found that reproduce the quantum mechanical predictions while being 
``local'' in this weak sense. In a way, it is a pity that the same word should thus be used 
to define two different concepts, having very different consequences. And, incidentally, 
it may be pointed out that the idea the mere outcome of a measurement should directly influence 
the outcome of another one, spatially separated from the former, appreciably violates our 
intuitive view on locality. So that, when all is said and done, Bell's definition appears as 
being distinctly more adequate than the ``weak'' one. At any rate, it is more in continuity with 
our commonsense understanding of the word. But it is true nevertheless that both do capture 
the same, reasonable, basic idea. To repeat, this idea intuitively is that, when two sequences 
S$_{\mbox{A}}$ and S$_{\mbox{B}}$ of events take place in two spatially separated space-time 
regions R$_{\mbox{A}}$ and R$_{\mbox{B}}$, 
the only instances in which S$_{\mbox{A}}$ and S$_{\mbox{B}}$ 
may happen to be correlated is when, within each 
sequence, the changes from one event to the next one are influenced by changes having 
taken place within some sequence S$_{\mbox{C}}$ of events pertaining to the common past of 
R$_{\mbox{A}}$ and R$_{\mbox{B}}$. 
So that, if subsequences S'$_{\mbox{C}}$ are considered in which the latter changes do not occur, 
the 
corresponding subsequences S'$_{\mbox{A}}$ and S'$_{\mbox{B}}$ should not be correlated. 
For further reference, 
let this condition be called ``{\it Condition C}''.
	
	For the purpose of avoiding an unreasonable dilution of the meaning of the word ``locality'' 
an appropriate convention is to restrict its use to the description of situations in which 
condition C is fulfilled. As we shall see, however, this condition is {\it not} fulfilled by the 
definition of locality adopted by Unnikrishnan.

\vspace{2mm}
\noindent{\it Remark}

	Unnikrishnan is roughly aware of what constitutes the main idea in condition C : the idea 
that an appropriately defined conditional joint probability of events A and B is to be written 
as a product. He calls it the ``separability of probability'' idea. Unfortunately, he does not seem 
to have grasped the crucial point concerning condition C, which is, as just explained, that it is 
only the probability concerning the subsequences S'$_{\mbox{A}}$ and S'$_{\mbox{B}}$ -
 otherwise said, the conditional 
probability of events corresponding to fixed {\bf h} values - that is expected to factorize. In 
Section 1.5 of [2] he describes an experiment in classical optics - the Hanbury Brown-Twiss 
experiment - where all physical aspects obey locality, and even the Bell inequality, but in 
which, he writes, the separability of probability is invalid. He considers this as showing 
that separation of probability is not a good criterion for locality, and, without explicitely 
saying so, he sort of suggests that this might be another flaw in Bell's proof. But in fact 
the argument misses its point. The Brown-Twiss experiment can be fully described within 
classical physics, that is, as a purely deterministic phenomenon in which every single 
detection event (call it A or B according to whether it takes place in one or in the 
other detector) is entirely determined by (a huge number of) elementary parameters that
 may be collectively designated by the symbol {\bf h}. Now, Unnikrishnan points out that the 
joint probability is not a product of the local probabilities at the two detectors and 
that there is a correlation of the intensities. But this is just because the joint 
probability he considers is the one that is experimentally measurable: which means 
that it is grounded on the consideration of a large sequence of elementary events, not 
all identical with one another, taking place at the source. In other words, the observed 
correlation merely reflects the fact that the sequences of events in the two detectors 
are due to common causes. Clearly, this joint probability has nothing to do with the one - 
call it, say, p' - that, in the proof of Bell's theorem, is considered as being a product. 
And indeed there is no reason whatsoever to assume that p' (which of course is not accessible 
to experiment) is {\it not} a product. Otherwise said, there is no reason to assume that the 
observed correlations are not due to common causes at the source, distributed according 
to the density $\rho$({\bf h}), In fact, here as in all deterministic theories, p' (roughly speaking 
and to within normalization) should be a product of two delta functions so that A and 
B are in fact two functions of h. The experimentally accessible correlation function 
can then be considered as having the form of Eq.(5) of [2], which, of course implies 
the validity of the Bell inequality.

\vspace{5mm}

\noindent{\bf 5 - A criticism of ``Unnikrishnan locality''}

	In Unnikrishnan's views a central role is played by the amplitudes C$_1$ and C$_2$ he 
introduces and by an associated ``amplitude correlation function'' U({\bf a},{\bf b}) =
 Real(NC$_i$C$_j^*$) where 
N is a normalization constant. And he claims that, in the model he puts forward, ``the 
locality assumption is strictly enforced since the two amplitudes C$_1$ and C$_2$ depend only 
on the local variables and on an internal variable [the difference between $\phi_1$ and 
$\phi_2$] 
generated at the source and then individually carried by the particles''. This may be 
considered as his (indirect) precise definition of what he means by ``locality''. It is 
easy to see that, as could be expected, it does not fulfill the above stated Condition C 
(which is essentially, as may be remembered, Bell's condition for locality). In the model 
in question consider, for example, the conditional probability P(+,+$\vert\phi_1$,$\phi_2$) 
that the 
outcomes of the measurements A and B should both be +1, under the condition that the phases 
$\phi_1$ and $\phi_2$ have fixed, given values. Since, in this model, $\phi_1$ and $\phi_2$ 
are the only parameters 
attached to the particles, they alone can be the carriers of the influences exerted by the 
source on the measuring instruments: so that if, within the sequence of emission events at 
the source, we consider a subsequence of such events within which the values of $\phi_1$ and $\phi_2$ do 
not vary from one event to the next one, according to condition C, the corresponding 
subsequences of A and B measurements outcomes should be uncorrelated. Which means that 
P(+,+$\vert\phi_1$,$\phi_2$) should be a product of two factors, one depending on
 $\phi_1$ and not on $\phi_2$ 
and the other one depending on $\phi_2$ and not on $\phi_1$. This, however, is not 
the case since, in 
the said model - according to assumption (3) of [2], Section 2.1 - P(+,+$\vert\phi_1$,$\phi_2$) 
is 
proportional (to within a numerical normalization factor) to the square of Re C$_{1+}$C$^*_{2+}$,
 that 
is to:
$$\cos^2[s(q_1 - q_2) +s(\phi_1 - \phi_2)].\ \ \  \mbox{(II)}$$
which is not such a product. 

	Now, admittedly anybody is free to use the words of the language the way he chooses provided 
he specifies the meaning he imparts to them. But, again, unrestricted use of such a freedom 
would obviously spread confusion. To avoid it, the best is, when we have to do with different 
concepts, to designate them by different names. In that spirit, the concept defined by 
Unnikrishnan, which is quite different from both Bell's and ``weak'' locality, could be called, 
say, ``Unnikrishnan locality''. It is a formal concept whose domain of meaningfulness seems to 
be restricted to the theories imparting to waves some type of ontological significance. With 
the help of the name we just gave to it we can state precisely what Unnikrishnan achieved and 
did not achieve. Actually what he has shown is that there is no proof that Unnikrishnan 
locality is violated in Nature. He did not show anything more. As we saw, he radically 
misunderstood the nature, significance and generality of Bell's theorem, and, quite definitely, 
he proved neither that it is false nor that it is irrelevant.

\vspace{5mm}
\noindent{\bf  6 - Conclusion}
 
	Unnikrishnan's meritorious investigations are interesting, not only in themselves but also 
in that they yield quite a vivid example of the multifarious difficulties and pitfalls that 
threaten the scientists within the field of research he elected. What comes out from the above
 is that:
	
1 - Unnikrishnan's interest for the detailed structure of quantum mechanics, and, in particular,
 state reduction at a distance may well be justified. Here, this problem was not investigated 
and consequently the possibility was not ruled out that, as Unnikrishnan suggests, 
self-consistency problems should still remain in this domain.

2 - Although several definitions of locality were put forward in the literature, they 
are not all equally satisfactory. They are all the more so as they are less arbitrary, 
that is, as they are closer to the everyday meaning of the word. In this respect, the 
Bell-like definition, also called by him ``local causality'', is, by far, the best one. 
In contrast, Unnikrishnan's definition is an abstract one and depends too much on the 
idea that wavefunctions have ontological reality. This makes it artificial.

3 - Last but not least: Unnikrishnan completely missed the substance and bearings of 
Bell's theorem, his assertions concerning its premises being  erroneous. In fact, the class of theories that this theorem rules out is
much wider than he considers it to be. Contrary to what he seems to suggest,
it includes any attempt at a realist local interpretation {\it of standard
quantum mechanics}.

\vspace{5mm}

\noindent{\bf References}

1 C.S.Unnikrishnan, {\it Resolution of the nonlocality puzzle in the EPR paradox},  Current Science 
{\bf 79}, 195 (200).

2 C.S.Unnikrishnan, {\it Is the quantum mechanical description of physical reality complete? 
Proposed resolution of the EPR puzzle},  Foundations of Physics Letters, {\bf 15}, N$^\circ$ 1,
 February 2002.

3 C.S.Unnikrishnan,  {\it Proof of Absence of Non-local State-reduction in Quantum Mechanics},  
preprint.

4 C.S.Unnikrishnan, {\it Einstein was right: Proof of absence of spooky state reduction in 
quantum mechanics}, http://arXiv,
quant-ph/0206175.

\end{document}